\documentclass[a4paper,12pt]{article}
\usepackage{float}
\usepackage{tabularx}
\usepackage{amsmath}
\usepackage{amsfonts}
\usepackage{adjustbox}
\usepackage{multirow}
\usepackage{color}
\usepackage{longtable}
\usepackage{pdflscape}
\usepackage{rotating}
\usepackage{xcolor}
\usepackage{xcolor}
\usepackage[colorlinks = true,
            linkcolor = blue,
            urlcolor  = blue,
            citecolor = blue,
            anchorcolor = blue]{hyperref}
\usepackage[margin=1in,footskip=0.25in]{geometry}
\title{Effective Range Approximation in Variable Phase Approach for Triplet $^3S_1^{\{np\}}$ and Singlet $^1S_0^{ \{nn, np, pp\}}$ State}
\author{Anil Khachi$^{1}$\\
$^{1}$ Department of Physics\\ St. Bedes College, 171002, \\Himachal Pradesh, India}
\begin{document}
\maketitle
\abstract{\noindent This work is a short communication where phase function method has been applied to obtain the phase shifts using Effective Range Approximation potential for $^3S_1-np$, $^1S_0-nn$, $^1S_0-np$, and $^1S_0-pp$ states. No free fitting parameters are used in calculations and reasonably good match with the experimental phase shifts is observed for E $\leq$ 20 MeV. Potentials are obtained for n-n, n-p and p-p scattering that are exponential well shaped.}
\\
\noindent {\textbf{keywords:} Variable phase approach, Effective range approximation, NN interactions}
\maketitle
\section{Introduction}
NN interaction is the most fundamental problem in nuclear physics which is yet not fully understood. Nuclear physicsts demands precision potentials that has a capability to obtain the properties of the interacting systems as well has predictive power. After Yukawa \cite{Yukawa} oldest theory of NN interactions which was based on mesons exchange, certain high class potentials were introduced like: Av-18, Nijm-I,II and III, CD-Bonn, Paris-Group, Hamada-Johnston, Yale-Group, Urbana-Group and Sao Paulo-Group CHPT potentials are having the capability to fit experimental data for other lighter systems \cite{Naghdi}. Recently Granada group \cite{granada2016} has performed partial wave analysis on 6713 published np and pp scattering data below 350 MeV (from 1950-2013), thus helped to construct the inverse potentials more accurately. Aoki \textit{et al.} have extracted the NN, hyperon potentials and the meson-baryon potentials potentials using lattice QCD.

The experimental setup in neutron-neutron scattering \cite{nn} consists of a neutron beam generated by $^3H(d,n)^4He$ reaction. The target consists of heavy water enriched in $D_2O$. In total the nd-breakup reaction i.e., $n+d(deuterium) \rightarrow p+n+n$ is the reaction leading to the appearance of interacting two neutrons in the final state. In addition to n-d, another reaction i.e., $\pi^+$ $\rightarrow$ $\gamma+n+n$ helps in determinining the scattering length for \textit{nn} reaction. For various NN intercations table \ref{table} presents different reactions that produces the desired projectile and different targets used for the same.\\
\begin{table}[]
\caption{Table represents various NN interactions taken up in this paper and different reactions responsible for producing desired projectiles. Last column presents the targets used in the scattering processes.}
\centering
\begin{tabular}{c|c|c|c}
\hline
Reaction & Projectile Production by & Projectile & Target            \\ \hline
n-n \cite{nn}       & $^{3}$H(d,n)$^{4}$He    & n          & D$_2$O               \\ 
n-p \cite{np}       & $^2$H(p,n)2p     & n          & H in CH$_2$               \\ 
p-p \cite{pp}       &   Ionizing hydrogen gas  & p          &  H in CH$_2$, W , -(C$_2$H$_4$)$_n$- or graphite \\ \hline
\end{tabular}
\label{table}
\end{table}

Neutron-neutron interaction is important because the reaction helps in the investigation of charge idependence which has been fundamentally interesting to the nuclear community. It is well known that to investigate the charge independence in nuclear forces it is quite sufficient to compare the \textit{np} and \textit{pp} interactions. It is also known that to get more accurate information about these forces one needs to get precise low energy scattering data. Under such circumstances only S-waves are particularly important. The investigation of low energy scattering parameters is directly related to test the hypothesis of charge independence and charge symmetry. Both theoreticians and experimentalists are keen to precisly determine the low energy parameters for \textit{np, pp} and \textit{nn} interactions.

The nucleon-nucleon interactions at low energies are usually expressend in terms of effective range ($r_e$) and scattering length (\textit{a}). These parameters provides information on the charge idependence of the nuclear forces. There have been various arguments regarding the possibility of violation of charge independence and charge symmentry and some of them are discussed here: 

\begin{itemize}
\item Darewych \textit{et al.} \cite{1} found that there was substantial difference between the \textit{np} and \textit{nn} potential curves with $^1S_0-nn$ ($V_0$=40.38 MeV) and for $^1S_0-np$ ($V_0$=61.99 MeV) hence observed ``violation of charge independence". The violation was credited to the unavailability of high energy \textit{nn} data. They used high energy \textit{np} and \textit{pp} data for fitting the \textit{nn} SPS curve.
\item Henley and others \cite{2} have discussed about the small departure from charge symmetry and charge indpendence. The departure can be originated by \textit{em} forces and is related to the basic understanding nuclear interaction and is ultimately connected with the fundamental particle principles of elementary physics. Since departures has been observed to be very small it was imperative to acquire precise data on the nucleon nucleon interaction.
\item Babenko and Petrov \cite{babenko} observed from a comparison of the low-energy parameters for the \textit{np} system and their counterparts \textit{pp} and \textit{nn} systems that the charge dependence of nuclear forces is violated, which is associated with the mass difference between the charged and neutral pions.
\item The values of scattering length and effective range are sensitive to any small change in n-n potentials. As stated by E.S. Konobeevski \textit{et al.} \cite{4}: ``\textit{A change of 7\% in V(r) may lead to change of 20-30\% in the scattering length calculation}".
\end{itemize} 
The above mentioned points are either related to the precise knowledge of data or related to the fundamental knowledge of elementary physics. Experimentally it has been observed that the p-p scattering \cite{pp} is even qualitatively very different from the n-p scattering. Because the Pauli principle excludes triplet states of even orbital angular momentum and singlet states of odd angular momentum. From the p-p scattering it is not possible to conclude directly that the n-p and p-p interactions are different. However there are arguments which make it seem fairly plausible that these interactions are indeed different.
Our main aim in this paper is to use VPA and to look for energy range where ERA potential is able to fit the experimental SPS data. In our earlier work we have applied VPA or PFM approach for various scattering systems like n-n, n-p, $n-\alpha$, n-d, $\alpha-\alpha$ and astrophysically important reaction $\alpha-^3He$ scattering systems  \cite{arxiv_nn}\cite{PRC}\cite{lalit}\cite{shikhaand}\cite{Khachialpha}\cite{Scripta}\cite{Brazilian} \cite{alpha_alpha_chitkara}\cite{alpha3He} where Variational Monte-Carlo methods have been employed for the optimization process \cite{aditi_chem}\cite{aditi_ajp}\cite{swapna_ejp}. We have also obtained n-p quantum mechanical functions like amplitude and wavefunctions for various n-p channels and are in good aggreement to that of Av-18 potential \cite{arxiv_np}.
\section{Methodology:}
We know from effective range approximation:
\begin{equation}
    k\cot(\delta)=-\frac{1}{a}+\frac{1}{2}kr_e^2
    \label{kcot}
\end{equation}
where \textit{a} and $r_e$ are scattering length  and effective range. If use is made of the approximation specified by Eq. \ref{kcot}, the S matrix can be written in the form \cite{babenko}:
\begin{equation}
    S(k)=\left(\frac{k+i\alpha}{k-i\alpha}\right) \left(\frac{k+i\beta}{k-i\beta}\right)
    \label{Smatrix}
\end{equation}
Where S(k) is related to scattering length \textit{a} and effective range $r_e$ by following relations
\begin{equation}
    \alpha=\frac{1}{r_e}\bigg[1-\big(1-\frac{2r_e}{a}\big)^{1/2}\bigg]
    \label{Alpha}
\end{equation}
\begin{equation}
    \beta=\frac{1}{r_e}\bigg[1+\big(1-\frac{2r_e}{a}\big)^{1/2}\bigg]
\end{equation}
The `velocity independent' or `static' potential corresponding to effective range approximation given by \cite{babenko}
\begin{equation}
\boxed{
V_{NN}(r) =-V_{NN0}\frac{e^{-r/R_{NN}}}{(1+\gamma e^{-r/R_{NN}})^2}}
\label{eq1}
\end{equation} 
where
\begin{equation}
R_{NN}=\frac{1}{2\beta}
\end{equation}
\begin{equation}
V_{NN0}=\frac{\hbar^2}{m_{NN}}\frac{\gamma}{R^2_{NN}}
\end{equation}
\begin{equation}
\gamma=\frac{1+2\alpha R_{NN}}{1-2\alpha R_{NN}}
\end{equation}
In above equations $m_{NN}$ is the reduced mass of nucleonic system nn, np and pp. The effective potential in equation \ref{eq1} will be introduced inside Variable phase approach equation as discussed below.
\subsection{Variable Phase Approach:}  
%%%%%%%%%%%%%%%%%%%%%%%%%%%%%%%%%%%%%%
The Schr$\ddot{o}$dinger wave equation for a spinless particle with energy E and orbital angular momentum $\ell$ undergoing scattering is given by
\begin{equation}
\frac{\hbar^2}{2\mu} \bigg[\frac{d^2}{dr^2}+\big(k^2-\ell(\ell+1)/r^2\big)\bigg]u_{\ell}(k,r)=V(r)u_{\ell}(k,r)
\label{Scheq}
\end{equation}
Where $k=\sqrt{E/(\hbar^2/2\mu)}$. Second order differential equation  Eq.\ref{Scheq} has been transformed to the first order non-homogeneous differential equation of Riccati type \cite{calogero}\cite{babikov} given by following equation:  
\begin{equation}
\delta_{\ell}'(k,r)=-\frac{V(r)}{k}\bigg[\cos(\delta_\ell(k,r))\hat{j}_{\ell}(kr)-\sin(\delta_\ell(k,r))\hat{\eta}_{\ell}(kr)\bigg]^2
\label{PFMeqn}
\end{equation}
Here in Eq. \ref{PFMeqn} prime denotes differentiation of phase shift with respect to distance and the Riccati Hankel function of first kind is related to $\hat{j_{\ell}}(kr)$ and $\hat{\eta_{\ell}}(kr)$ by $\hat{h}_{\ell}(r)=-\hat{\eta}_{\ell}(r)+\textit{i}~ \hat{j}_{\ell}(r)$ . In integral form the above equation can be simply written as:
\begin{equation}
\delta(k,r)=\frac{-1}{k}\int_{0}^{r}{V(r)}\bigg[\cos(\delta_{\ell}(k,r))\hat{j_{\ell}}(kr)-\sin(\delta_{\ell}(k,r))\hat{\eta_{\ell}}(kr)\bigg]^2 dr
\end{equation}
Eq.\ref{PFMeqn} is numerically solved using Runge-Kutta 5$^{th}$ order (RK-5) method with initial condition $\delta_{\ell}(0) = 0$. For $\ell = 0$, the Riccati-Bessel and Riccati-Neumann functions $\hat{j}_0$ and $\hat{\eta}_0$ get simplified as $\sin(kr)$ and $-\cos(kr)$, so Eq.\ref{PFMeqn}, for $\ell = 0$ becomes 
\begin{equation}
\delta'_0(k,r)=-\frac{V(r)}{k}\sin^2[kr+\delta_0(k,r)]
\end{equation}
%%%%%%%%%%%%%%%%%%%%%%%%%%%%%%%%%%%%%%%%%%%%%%%%%
%In above equations $k=\sqrt{E/(\hbar^2/2\mu)}$ and $\hbar^2/2\mu$ = 41.47 MeV fm$^{2}$ for np case.
In above equation the function $\delta_0(k,r)$ was termed ``Phase function'' by Morse and Allis \cite{morse}. The significant advantage of \textit{PFM} method is that, the phase-shifts are directly expressed in terms of the potential and have no relation to the wavefunction. This has been utilised in this paper to obtain inverse potentials in an innovative way by implementing a modified \textit{VMC} in tandem with \textit{PFM}. The technique optimizes the model parameters of the potential to obtain the best match with respect to the experimental SPS values. Also, rather than solving the second order Schr$\ddot{o}$dinger equation, we only need to solve the first order non-homogeneous differential equation whose asymptotic value gives directly the value of \textit{SPS}.
\section{Results and Discussion}
\begin{figure*}
\centering
{\includegraphics[scale=0.5,angle=0]{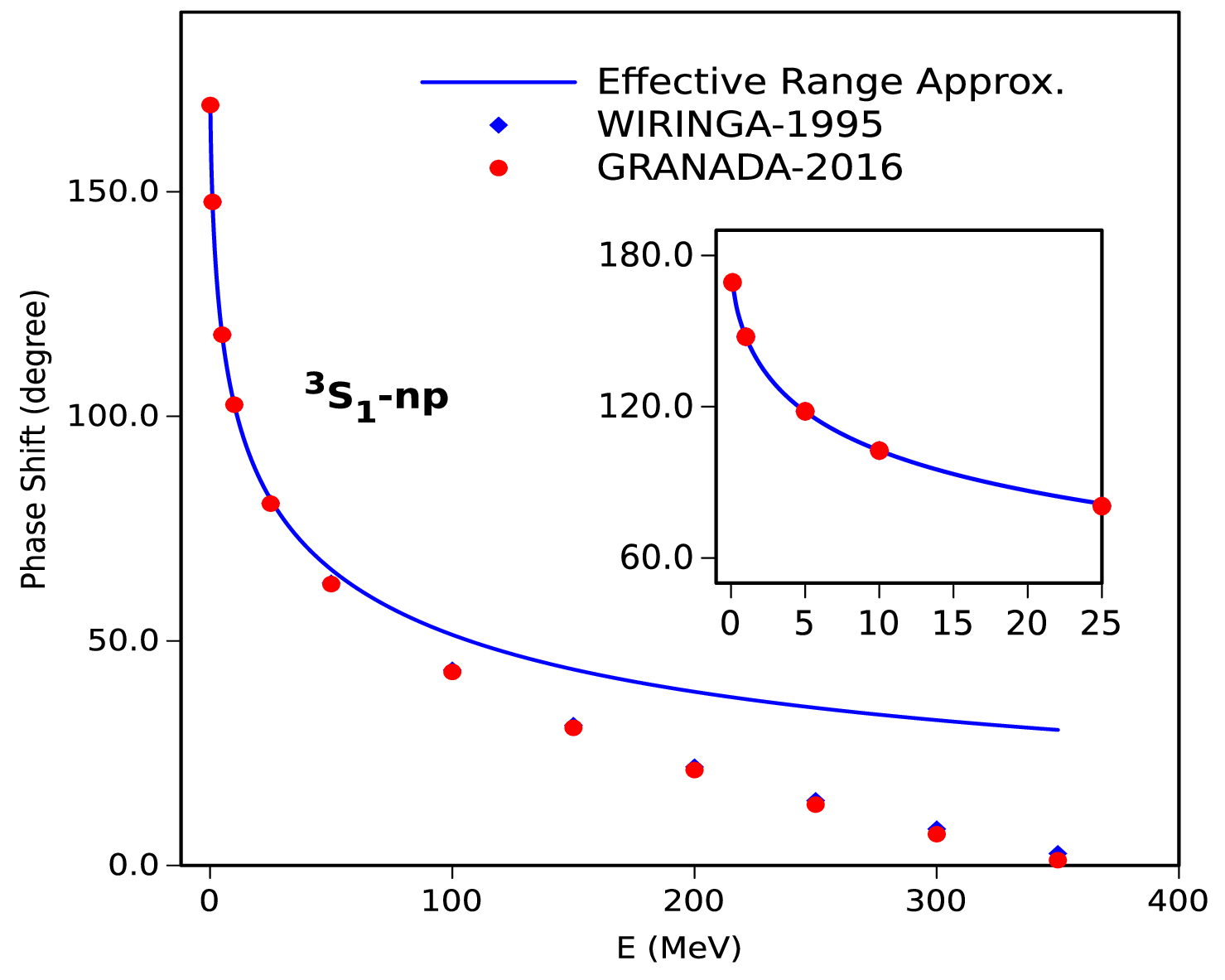}}
\caption{Bound triplet state $^3S_1-np$ SPS variation in comparison to experimental data from WIRINGA-1995 \cite{wiringa} and GRANADA-2016 \cite{granada2016}.}
\label{SPS_POT1}
\end{figure*}

\begin{figure*}
\centering
{\includegraphics[scale=0.5,angle=0]{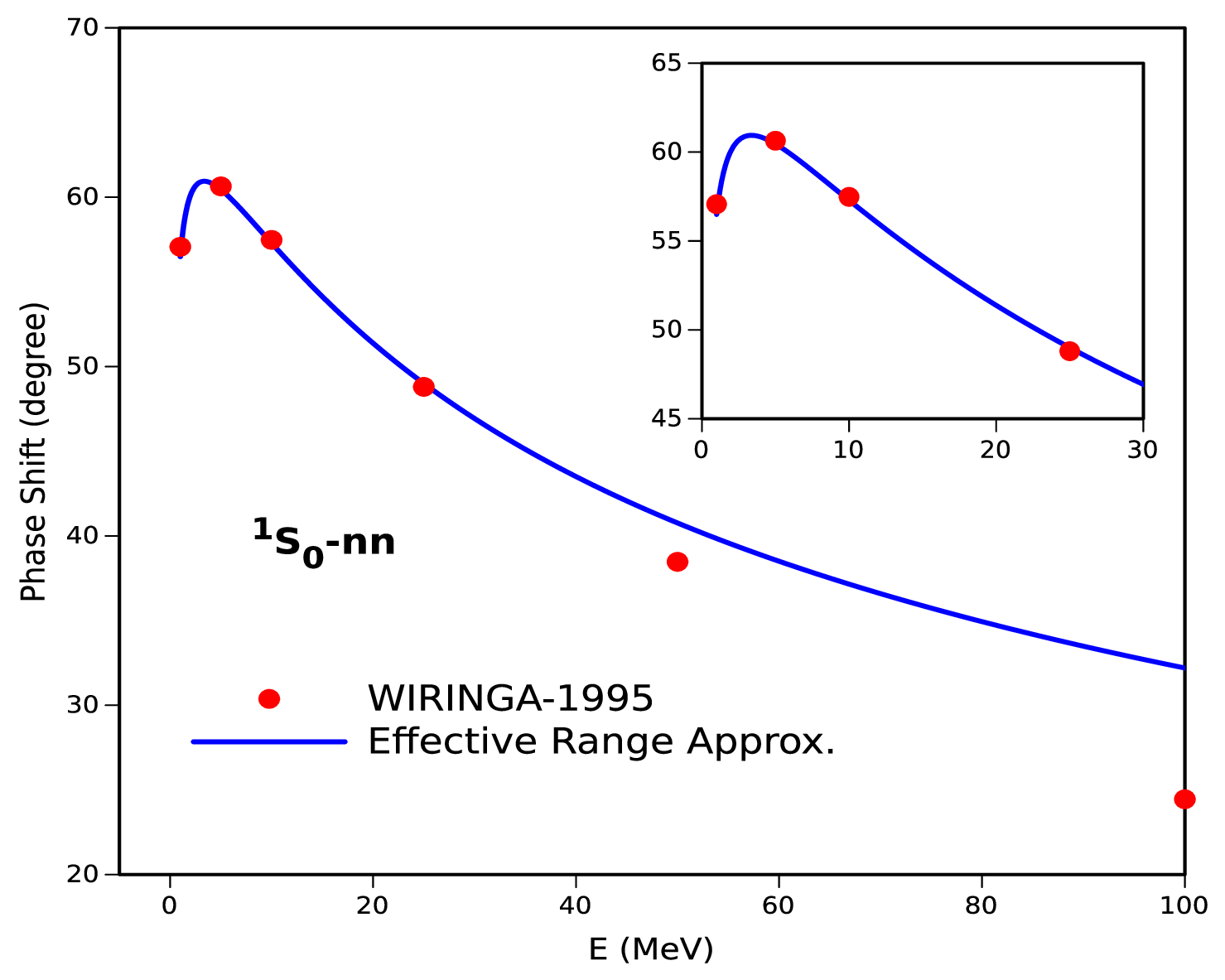} }
\caption{Singlet state $^1S_0-nn$ SPS variation in comparison to experimental data from WIRINGA-1995 \cite{wiringa}.}
\label{SPS_POT2}
\end{figure*}

\begin{figure*}
\centering
{\includegraphics[scale=0.45,angle=0]{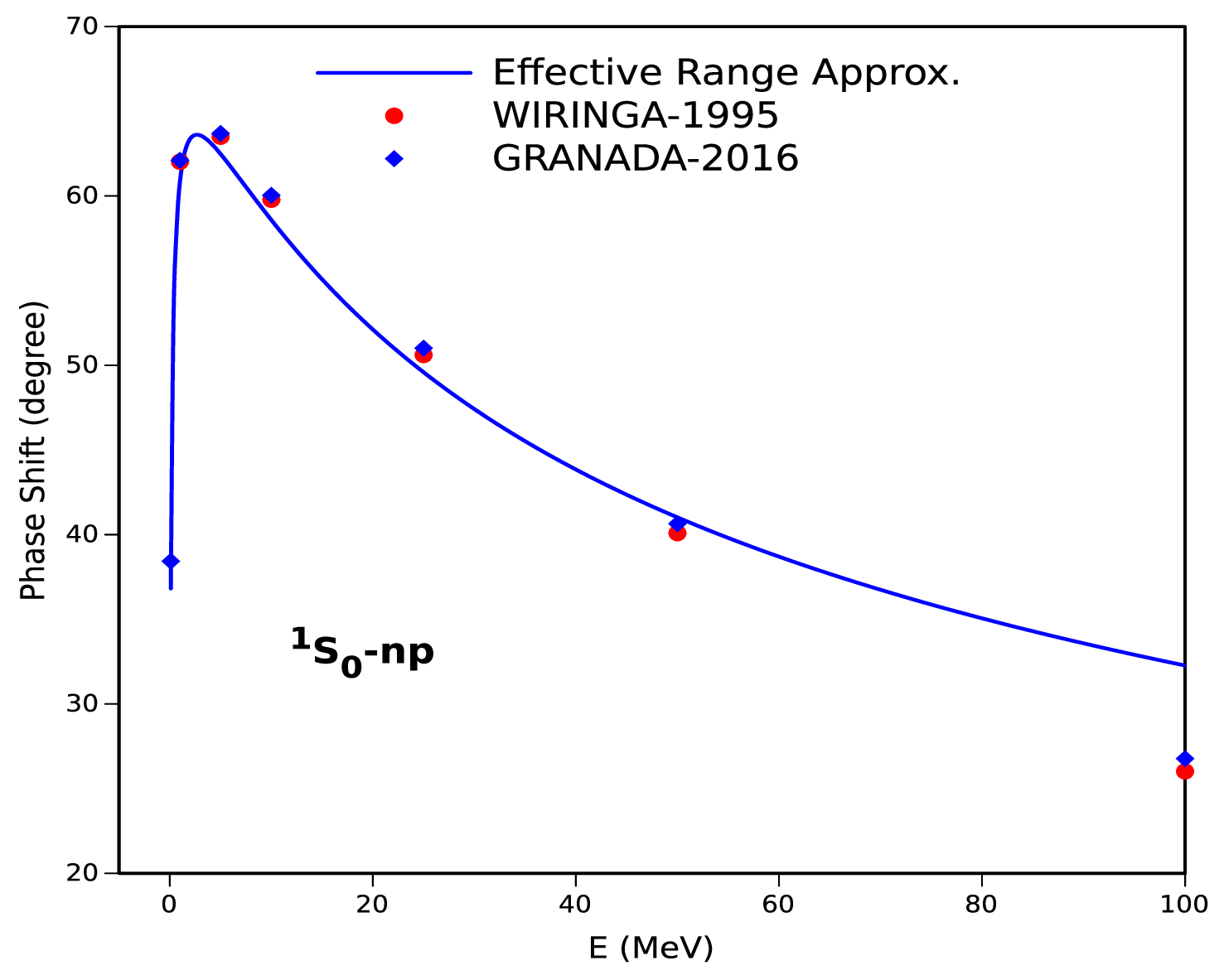} }
\caption{Singlet state $^1S_0-np$ SPS variation in comparison to experimental data from WIRINGA-1995  \cite{wiringa} and GRANADA-2016 \cite{granada2016}.}
\label{SPS_POT3}
\end{figure*}

\begin{figure*}
\centering
{\includegraphics[scale=0.8,angle=0]{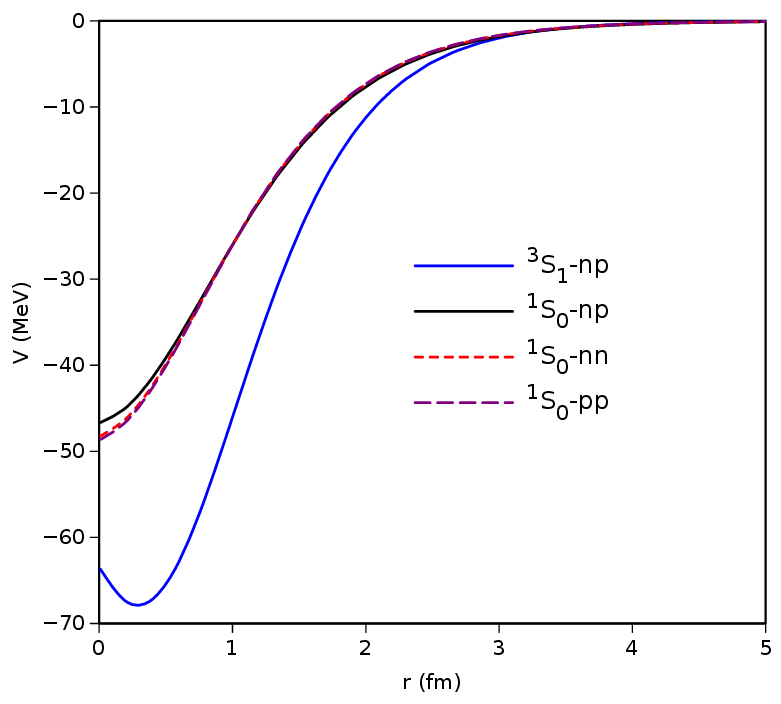} }
\caption{Potentials for $^3S_1$-np and $^1S0$-np, nn and pp scattering systems.}
\label{SPS_POT5}
\end{figure*}

\begin{itemize}
\item For $^3S_1$ state: We can see from figure \ref{SPS_POT1} that the SPS are in excellent match with analysis data of both Wiringa \textit{et al.} \cite{wiringa} and  Granada group \cite{granada2016} upto 20 MeV of projectile energy in laboratory frame. The corresponding potential for $^3S_1$ state is shown in figure \ref{SPS_POT5}, where depth of around 70 MeV is observed.
\item For $^1S_0$-nn and np again the SPS are in good match with Wiringa \textit{et al.} and  Granada as shown in figure \ref{SPS_POT2} and \ref{SPS_POT3}. After 20 MeV the deviations can be clearly seen in figure \ref{SPS_POT2} and \ref{SPS_POT3} indicating the absence of repulsive core which becomes important at higher energies. The potentials are shown in figure \ref{SPS_POT5}.
\item Finally for $^1S_0$-pp state the SPS are in reasonably accuracy with Wiringa \textit{et al.} and  Granada group. The combined potentials for all $^1S_0$-nn, np and pp state can be seen in \ref{SPS_POT5} to be very tightly packed with each other indicating the charge independence hypothesis of nuclear force.
\end{itemize}
It is observed that all the SPS are positive and their corresponding potentials are negative indicating that positive  phase shifts will have attractive interactions which is in accordance to the main VPA equation. This point is of great pedagogic value as we can just by observing the nature ($+/-$) of SPS can predict the nature of the interaction potential ($-/+$). But above statement is not stringent as the results may change for potential having both attractive part and repulsive core as well as for higher values of wavenumber k.  
\begin{figure*}
\centering
{\includegraphics[scale=0.45,angle=0]{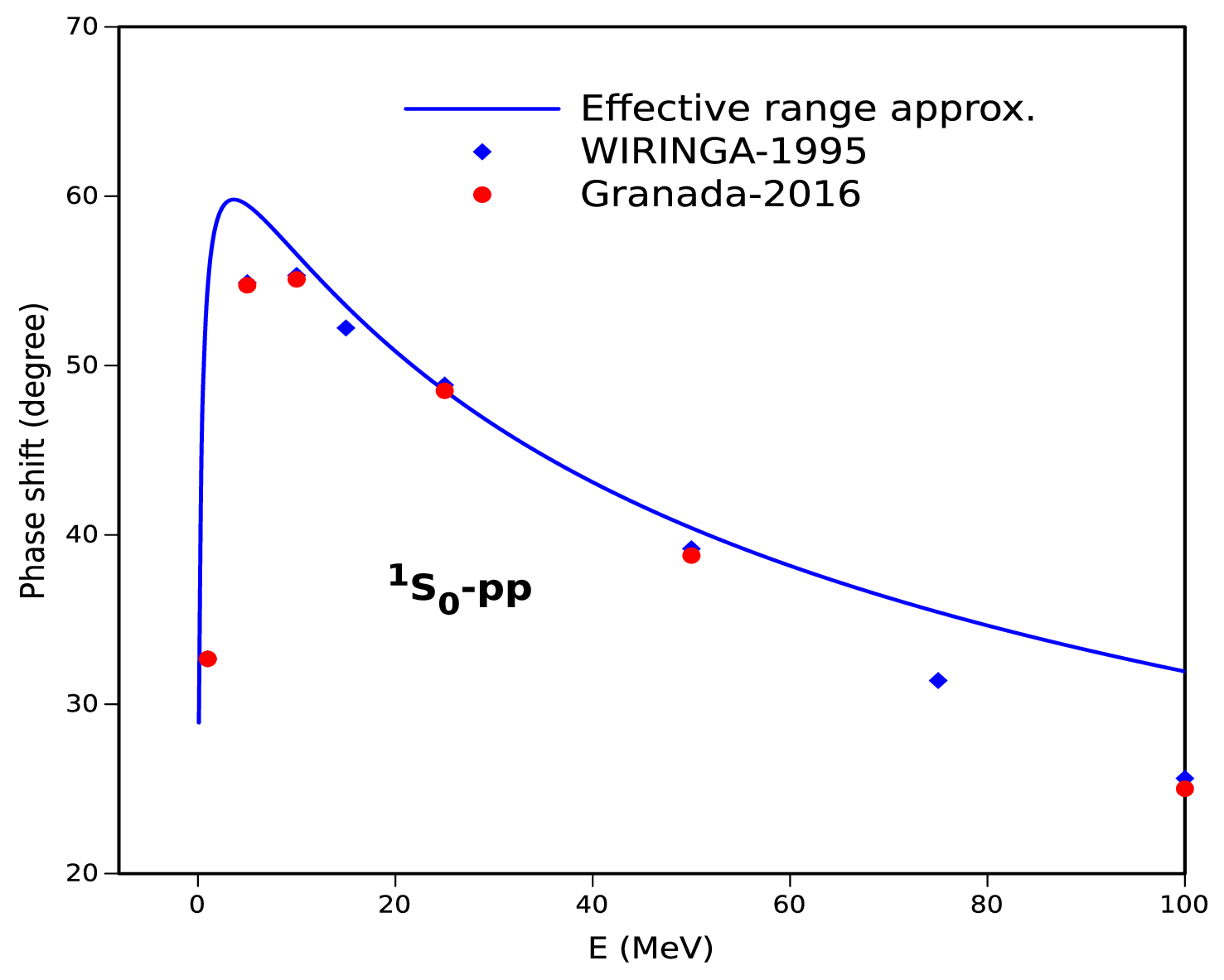} }
\caption{Singlet state $^1S_0-pp$ SPS variation in comparison to experimental data from WIRINGA-1995  \cite{wiringa} and GRANADA-2016 \cite{granada2016}.}
\label{SPS_POT4}
\end{figure*}

\section{Conclusion}
Using ERA in VPA we obtained the scattering phase shifts for n-n, n-p and p-p scattering. Good match in SPS with experiment is found for energy upto 20 MeV after which deviations are seen for all $^1S_0$-(np, nn \& pp) states. The reason for deviation is that at very small inter-nucleon distances, a strong repulsive core is expected due to strong anti-corelation between nucleons hence the absence resulting into deviations observed. As stated by Breit \textit{et al.}: ``\textit{Any of potential shapes like Square well, Gaussian well, exponential well and yukawa well would give satisfactory agreement with the experimental data, provided the range and depth of the well are suitably chosen}". The presented potential is also exponential shaped and good enough to fit low energy SPS just by giving scattering parameters as inputs within ERA potential.

\section{Acknowledgements} This work is devoted to Prof. Francesco Calogero and my Ph.D supervisor Prof. O. S. K. S. Sastri.

\end{document}